\documentclass[11pt]{article}
\usepackage{amssymb}

\newcommand{\bi}{\bibitem}
\newcommand{\be}{\begin{eqnarray}}
\newcommand{\ee}{\end{eqnarray}}
\newcommand{\rar}{\rightarrow}

\begin{document}

\begin{titlepage}

\title{Antimatter in the Milky Way}

\author{C. Bambi$^{a,b,}$\footnote{E-mail: bambi@fe.infn.it} and
A.D. Dolgov$^{a,b,c,}$\footnote{E-mail: dolgov@fe.infn.it}}

\maketitle

\begin{center}
$^{a}$Istituto Nazionale di Fisica Nucleare, Sezione di Ferrara,
       I-44100 Ferrara, Italy\\
$^{b}$Dipartimento di Fisica, Universit\`a degli Studi di Ferrara,
       I-44100 Ferrara, Italy\\
$^{c}$Institute of Theoretical and Experimental Physics,
       113259, Moscow, Russia
\end{center}

\vspace{0.5cm}

\begin{abstract}
Observational signatures of existence of antimatter objects 
in the Galaxy are discussed. We focus on point-like sources 
of gamma radiation, diffuse galactic gamma ray background and 
anti-nuclei in cosmic rays.
\end{abstract}

\end{titlepage}

\section{Introduction}

One can conclude from simple considerations that 
there is much more matter than antimatter 
around us~\cite{steigman, stecker}. 
The Earth and the Solar System are evidently made of matter 
and the very small antiproton-to-proton ratio in cosmic rays,  
$\sim 10^{-4}$, suggests a secondary origin of antiprotons 
by cosmic ray collisions in the interstellar medium and an absence of 
a large amount of antimatter in the Galaxy. 
Still an excess of cosmic antiprotons at low 
energies~\cite{low-anti-p} might point to non-standard 
sources of their production and, in particular, to some
antimatter objects in our neighborhood. A similar
conclusion may be obtained from observations of 
gamma rays originating from $e^+e^-$ 
annihilation \cite{e-continuum,e-line}. Though a conventional
mechanism of positron production is the most probable one,
light dark matter might be a possible source of positrons,
especially because of the observation of the 511 keV line from 
the halo emission, see the above quoted papers. 
This explanation suffers from a necessity to fine-tune
the mass of the dark matter particles, so that they would
decay or annihilate into non-relativistic positrons. Another 
possible source of positrons could be primordial antimatter 
in the Galaxy. For other galaxies the sensitivity is not
high enough to exclude or observe significant $\bar p p$ and 
$e^+ e^-$ annihilation indicating possible cosmic antimatter objects, 
but galaxies dominated by antimatter are excluded 
in our galactic cluster, i.e. 
up to distances of about 10 Mpc \cite{steigman}, 
by non-detection of $\gamma$ rays which would be 
produced by the annihilation of the galactic antimatter with 
the matter from the infalling intergalactic gas.

On the other hand, {\it a priori} we could expect an approximately 
charge symmetric universe, or at least a universe with some
considerable amount of antimatter, since matter and antimatter 
have (almost) the same properties.

Cosmological matter-antimatter asymmetry can be expressed
in terms of the baryon-to-photon ratio, $\beta$. The
Big Bang Nucleosynthesis (BBN) \cite{pdg} and the Cosmic 
Microwave Background Radiation (CMBR) \cite{wmap} provide 
two independent measurements which are consistent with
each other 
\be
\beta = \frac{n_B-n_{\bar B}}{n_\gamma} 
\approx 6 \cdot 10^{-10} 
\label{eta-obs}
\ee
where $n_B \gg n_{\bar B}$, whereas the freeze-out abundances 
in a homogeneous baryo-symmetric universe would be
$n_B/n_\gamma = n_{\bar B}/n_\gamma \sim 10^{-18}$ \cite{zeld65}.

There can be three possible types
of cosmological matter-antimatter asymmetry:
\begin{enumerate}
\item $\beta$ is constant and the universe is 
100\% matter dominated.
\item The universe is globally baryo-symmetric. It
consists of equal amount 
of similar domains of matter and antimatter.
\item The universe has a non-vanishing average baryonic charge, 
but $\beta$ is not spatially constant and can even be negative
in some space regions. In other words
there could be lumps of antimatter
in matter dominated universe. 
\end{enumerate}
The type of the asymmetry depends upon the mechanism 
of baryogenesis which took place in the early universe.
The baryogenesis scenarios mostly focus on the first 
possibility, for review see Refs.~\cite{ad1}-\cite{rubakov}, but at 
present there is neither experimental nor observational 
evidence in favor of one model 
over another, since the involved physics operated
at such high energies that it is difficult or impossible
to test it in laboratories on the Earth.
On the other hand, a globally baryo-symmetric universe is 
certainly theoretically appealing, but it seems observationally 
excluded or, to be more precise, the size of the domain where we 
live should be (at least) comparable to the present day 
cosmological horizon \cite{cdg}.

Most interesting phenomenologically is the third case
because it allows for existence of anti-objects in 
our neighborhood and hence 
for peculiar features which may be observed
in a near future, thanks to the advent of antimatter research
projects such as AMS and PAMELA. A small amount of antimatter
is not excluded even nearby in the Galaxy. The
aim of our paper is to consider phenomenological manifestation
of that. The effects from antimatter objects in the Galaxy
were analyzed also in Refs. \cite{early-khlop} for a different
mechanism of antimatter creation and because of that for
restricted types of such objects.

The content of the paper is the following. 
In Sec.~\ref{s-mechanisms} we briefly describe a 
mechanism for generation of lumps of antimatter 
in baryon dominated universe. Their cosmological evolution 
is considered in Sec.~\ref{s-evolution}
and matter-antimatter annihilation in contemporary universe
is discussed in Sec.~\ref{s-late}. 
In Sec.~\ref{s-sources}
we focus on point-like sources of $\gamma$ rays,
in Sec.~\ref{s-background} on the diffuse galactic gamma ray 
background, and in Sec.~\ref{s-cr} on the possibility of 
anti-nuclei in cosmic rays. In Sec.~\ref{s-spectacular}
we speculate on more violent events, where large amounts
of matter and antimatter come into contact.
The results are summarized in Sec.~\ref{s-conclusion}.

\section{Antimatter in baryon asymmetric universe}
\label{s-mechanisms}

Baryon asymmetric universe with high density regions
of matter and antimatter could be created if there existed
two different sources of CP violation \cite{ad-cp}: the 
background baryon asymmetry could be provided by an explicit 
violation of CP in the Lagrangian, whereas small bubbles 
with very high baryon asymmetry could be produced by the 
presence of a stochastic or dynamical violation of CP. 
The concrete models are considered e.g. in 
Refs.~\cite{ad-cp,anti-rev}.

In what follows we will use as a reference the scenario 
proposed in paper~\cite{ad-silk}.
The starting point is the Affleck-Dine mechanism \cite{a-d},
where scalar fields with non-zero baryonic and/or leptonic
charges (predicted in supersymmetric extensions of the
Standard Model) have the potential with flat directions, that 
is the directions along which the potential energy does not 
change. As a toy-model, we can consider the potential
of the form:
\be\label{potential}
U(\chi) = \lambda |\chi|^4 (1-\cos 4\theta) \; ,
\ee
where $\chi = |\chi| e^{i\theta}$ is the scalar field with
baryonic charge $B \neq 0$. Potential (\ref{potential})
has four flat directions along $\cos 4\theta=1$. It is not
invariant under the transformation $\chi \rar \chi e^{i\alpha}$,
i.e. $B$ is not conserved. Due to the infrared instability of
massless (or light, $m_\chi \ll H$) fields in de Sitter
space-time \cite{infrared} during 
inflation, $\chi$ can condense along the flat directions
of the potential, acquiring a large expectation value. 
In the course of the cosmological expansion the Hubble
parameter drops down and when
the mass of the field, $m_\chi$, exceeds the universe expansion
rate $H$, $\chi$ evolves to the equilibrium point $\chi=0$
and the baryonic charge stored in the condensate is transformed
into quarks by $B$-conserving processes. Since here CP is violated
stochastically by a chaotic phase of the field $\chi$, then
during the motion to the equilibrium state the
matter and antimatter domains in a globally symmetric universe
would be created. An interesting feature of the model is that 
regions with a very
high $\beta$, even close to one, could be formed.

If the scalar field $\chi$ is coupled to the inflaton $\Phi$
with an interaction term of the kind
\be
V(\chi,\Phi) = \lambda |\chi|^2 (\Phi - \Phi_1)^2 \; ,
\ee 
the ``gates'' to the flat directions might be open only for
a short time when the inflaton field $\Phi$ was close to $\Phi_1$.
In this case, the probability of the penetration to the flat 
directions is small and $\chi$ could acquire a large expectation
value only in a tiny fraction of space. The universe would have
a homogeneous background of baryon asymmetry 
$\beta \sim 6 \cdot 10^{-10}$ generated by the same field 
$\chi$ which did not penetrate to larger distance through the  
narrow gate or by another mechanism of baryogenesis, while the 
high-density matter, $\beta >0$, and antimatter, $\beta < 0$,
regions would be rare, while their contribution to
the cosmological mass density might be significant or even
dominant. Let us call these 
bubbles with high baryonic number density B-balls.

Since the conditions for B-ball creation had been prepared
during inflation, their initial radius, $R_B$, might quite
naturally exceed the cosmological horizon.

In a simple model, the mass spectrum of 
B-balls has a log-normal form \cite{ad-silk}
\be\label{dis}
\frac{dN}{dM} = C \exp 
\Big[-\gamma \ln^2 \Big(\frac{M}{M_0}\Big)\Big] \; ,
\ee
where $C$, $\gamma$ and $M_0$ are unknown parameters of the
underlying theory. Depending on their values and on $\beta$,
which are stochastic quantities, such bubbles could form 
clouds of matter or antimatter with high (anti)baryon
number density, more compact object like (anti)stars, and
primordial black holes. If all antimatter is hidden inside 
anti-black holes it would be unobservable. However, it
is natural to expect that there could be abundant 
anti-stars (either normal or compact ones, as e.g.
white dwarfs) or clouds of antimatter with higher than
normal baryon density. Such antimatter objects
could survive in the early universe due to their
higher density, invalidating the bounds of Ref.~\cite{early-khlop}.

The compact matter/antimatter objects created by the 
described mechanism might make a part of, or even the 
whole, cosmological dark matter\footnote{According to 
Ref.~\cite{cdm}, the present observational data allow (at the 
95\% C.L.) macroscopic compact objects with masses greater 
than $10^{-2} \, M_\odot$ to constitute 88\% or less of 
the cosmological dark matter, so 100\% is not 
highly probable but 
still allowed.}. An interesting feature of such dark 
matter is that it consists of (stellar size) ``particles'' 
with dispersed masses. In particular, even if $M_0$ is 
close to the solar mass, there is a non-zero probability
that on the tail of distribution (\ref{dis})
very heavy black holes of millions solar masses might be created.
In particular, if there is one heavy black hole per galaxy,
they could serve as seeds for galaxy formation. On the other
hand, in the non-collapsed regions with high baryonic
number density primordial nucleosynthesis
proceeded with large $\beta$, producing nuclei heavier than
those formed in the standard BBN~\cite{ibbn}. If these 
regions are in our neighborhood, an observation of heavy 
anti-nuclei in cosmic rays would be plausible.

A different model of creation of (much smaller size) compact
anti matter objects was suggested in Ref.~\cite{zhit1}. Their
observational signatures in cosmic gamma ray radiation 
are analyzed  in Ref.~\cite{zhit2}.

\section{Cosmological evolution of B-balls
\label{s-evolution}}

When baryogenesis was over, the regions with high values of $|\beta|$
had the same energy density as the background. Only the chemical content 
was different. These are the so called isocurvature density perturbations.
On the boundary of the bubbles the density contrast should be non-zero
due to the gradient term, $|\partial \chi|^2$, but it is relatively
insignificant. Moreover, this term disappeared at later stage when
$\chi$ relaxed down to the equilibrium point, $\chi = 0$. 
After the electroweak phase transition, when quarks acquired masses due to 
the Higgs field condensate, the density contrast became nonzero,
especially because of the large mass of $t$-quark, $m_t \sim 150$ GeV.
A short life-time of $t$ and other heavy quarks and a large number 
of different quark species in the 
primeval plasma make the density contrast relatively small and we ignore
it in what follows.

More essential is the QCD phase transition when quarks became confined
forming nonrelativistic nucleons. The density contrast of 
B-bubbles would be equal to:
\be
r_B =\frac{\delta \rho}{\rho} = \frac{\beta n_\gamma m_N}{\pi^2 g_* T^4/30}
= \beta\,\frac{0.7 m_N}{g_* T} \approx 0.068\,\beta\, 
\left(\frac{10.75}{g_*}\right)\,
\left(\frac{m_N}{T}\right) \, ,
\label{r-B}
\ee
where $m_N \approx 940 $ MeV is the nucleon mass, $T$ is the temperature
of the cosmic plasma, and $g_* \approx 10$ is the number of particle
species ``living'' in the plasma. The nonrelativistic
baryonic matter starts to dominate inside the bubble at 
\be
T =T_{in} \approx 65 \, \beta \,{\rm MeV} \, .
\label{T-eq}
\ee
The mass inside a baryon-rich bubble at the radiation dominated stage is 
\be
M_B = \frac{4\pi}{3}\,R_B^3 \rho = (1+r_B)\,M_{Pl}^2 t \,
\left(\frac{R_B}{2t}\right)^3 
\approx 2\cdot 10^5 \, M_\odot (1+r_B) \left(\frac{R_B}{2t}\right)^3\,
\left(\frac{t}{\rm sec}\right) \, ,
\label{M-B}
\ee
where $M_\odot$ is the solar mass and
we assumed $\rho = 3M_{Pl}^2(1+r_B)/32 \pi t^2$.
If at the moment of the horizon crossing,
i.e. at $R_B = 2t$, the density contrast would be large, 
$r_B \geq 1$, the bubble
gravitational radius $R_g$ would be larger than $R_B$ and 
the bubble should form a primordial black hole (BH).  
The BH masses are determined by 
the plasma temperature at the moment of horizon crossing.
Assuming the approximate relation 
$t/{\rm sec} \approx \left({\rm MeV}/T\right)^2$,
we find $M_{BH} \approx 4\cdot 10^5 \, M_\odot\, (t/$sec).
The cosmological QCD phase transition took place at 
$t = 10^{-5} - 10^{-4}$ sec and the masses of the first formed BHs should be
near 10 solar masses. The baryon-rich bubbles with larger radius could make
much heavier BHs. 

Smaller bubbles with $R_B<2t$ and $r_B \sim 1$ could form compact stellar
type objects with masses given by Eq. (\ref{M-B}) and with
the initial (that is at $T=T_{in}$) mass density of non-relativistic matter:
\be
\rho_B \approx 4.4\cdot 10^5\, r_B\,
\left(\frac{{\rm sec}}{t_{in}}\right)^2\, {\rm g/cm}^3 \approx
10^{13} \beta^4 \,\, {\rm g/cm}^3\, ,
\label{rho-B}
\ee
where Eq. (\ref{T-eq}) and $t/{\rm sec} \approx \left({\rm MeV}/T\right)^2$
were used.

The subsequent evolution of B-balls depends upon the ratio of their mass
to the Jeans mass.  
The Jeans wave length is: 
\be
\lambda_J = c_s \sqrt{\frac{\pi M_{Pl}^2}{\rho}} =
 t\,\sqrt{\frac{32 \pi^2}{3}\,\frac{T}{m_N}} 
\approx 10 t\,\sqrt{T/m_N} 
\label{lambda-J}
\ee
where the speed of sound is $c_s \approx \sqrt{T/m_N}$ and 
$\rho = 3m_{Pl}^2 /32\pi t^2$. It is evident from this expression that
$\lambda_J $ exceeds horizon, $l_h=2t$, if $T<40$ MeV. If we take
the time and temperature in Eq. (\ref{lambda-J})
at the beginning of matter dominance in the B-bubble given
by Eq. (\ref{T-eq}), the initial value of 
$\lambda_J$ would be 
\be
\lambda_J^{(in)} \approx 6.2\cdot 10^{-4} \,\beta^{-3/2} \,\,{\rm s} \, .
\label{lambda-in}
\ee
Correspondingly the initial value of the Jeans mass is
\be
M_J \approx 135\left({T}/{m_N}\right)^{3/2} M_{Pl}^2 t
\approx 100 M_\odot \beta^{-1/2} \, .
\label{M-J}
\ee
For relativistic expansion the temperature drops
as $T \sim 1/a$, where $a(t)$ is the cosmological scale factor.
In such a case the Jeans mass would rise with time. 
However, the temperature of matter dominated B-bubbles drops much
faster,  $T\sim 1/a^2$, and the Jeans mass would decrease with time
in agreement with intuitive expectations. According to Eq. (\ref{M-J})
Jeans mass scales as $M_J \sim T^{3/2} t$. For nonrelativistic expansion
law, $a \sim t^{2/3}$ and consequently 
\be
M_J \sim 100 M_\odot\, (a_{in}/a)^{3/2},
\label{M-J-low}
\ee
where $a_{in}$ is cosmological scale factor at the onset of matter
dominance inside B-ball.

$M_J$ became of the order of the solar mass for $a/a_{in} \sim 20$.
For an estimate let us take $T_{in}\sim 10$ MeV. According to
Eq. (\ref{T-eq}), it is realized for $\beta \approx 0.15$. Correspondingly
the temperature inside B-bubble when $M_J$ dropped down to $M_\odot$, 
would be 
\be
T_{in} (a_{in}/a)^2 \approx 0.025\,\,{\rm  MeV}. 
\label{T-msun}
\ee

The mass density
evolved as $1/a^3$ and thus it became
\be
\rho_B = \rho_B^{(in)} (a_{in}/a)^{3} \approx 6\cdot 10^5\,\,
{\rm g/cm^3}
\label{rho-Bb}
\ee
to the moment when $M_J=M_\odot$. Correspondingly the radius of such B-ball
(i.e. the Jeans wave length) would be 
$R_B \sim 10^9 $ cm.

It is noteworthy that B-bubbles would be supported against 
expansion by the pressure of the hotter surrounding relativistic 
plasma. Since the photon mean free path inside B-bubbles is quite 
small, see below, the heat exchange takes place only on the 
surface and is not efficient. The external pressure slows down the 
expansion of B-balls and their adiabatic cooling.

If $\lambda_J < R_B $, the baryon-rich bubbles with such radius would 
decouple from the cosmological expansion and for $R_B < 2t$ they
would not be black holes but compact star-like objects. According to 
Eq. (\ref{rho-Bb}) their mass density could be much larger than the
density of the normal stars and similar to that of compact stars.
They might survive to the present time or, if nuclear reactions inside
would be significant, they would explode enriching outer space with
heavy elements or anti-elements. Since their temperature at the moment
of formation is very high, the pressure is also high, higher than the
usual pressure in normal stars powered by nuclear reactions. So we 
assume that such objects mostly survive to the present time. After a 
while, their temperature would become larger than the cosmological 
temperature because they stopped expanding and the temperature is not
red-shifted. They would cool down because of radiation from the surface
into external colder space. Their luminosity 
can be estimated as $L \sim T^4_{surf} R_B^2$, where $T_{surf}$ is the
surface temperature. With infinitely large thermal conductivity,
the life-time with respect to the cooling would be quite short
$\tau \sim T^4 R_B^3 /L \sim R_B$. Of course this result is an underestimate
and the real cooling time must be much longer. Without additional sources
of energy, the cooling time should be of the order of the photon diffusion 
time, $t_{diff}$, inside these compact objects.  If the photon mean
free path $l_{free}$ is smaller than the radius $R_B$ the diffusion time
from the center to the surface can be estimated as
\be
t_{diff} \sim R_B^2/\tau_{free} \, .
\label{t-diff}
\ee
The mean free time of photons is equal to
\be
\tau_{free} \sim 1/(\sigma_{e\gamma} n_e) \, ,
\label{tau-free}
\ee
where $n_e$ is the number density of electrons and 
$\sigma_{e\gamma} $ is the photon-electron
cross-section. For low energy photons, $E_\gamma < m_e$, 
it is equal to the Thomson
cross-section, $\sigma_{Th} = 8\pi \alpha^2/ 3 m_e^2$. For large 
energies it drops as inverse center-of-mass energy squared.

The photon diffusion is quite slow and the cooling time of baryon-rich
compact objects is determined by $t_{diff} $. Assuming that
$\sigma_{e\gamma} = \sigma_{Th}$ we obtain:
\be
t_{diff} \approx 2\cdot 10^{11}\,{\rm sec} 
\left(\frac{M_B}{M_\odot}\right)\,\left(\frac{\rm sec}{R_B}\right) 
\left(\frac{\sigma_{e\gamma}}{\sigma_{Th}}\right)\, .
\label{tau-free-num}
\ee
The mass and size of the baryonic bubbles depends upon the parameters of
the model of their creation. We assume, for the sake of estimate, 
that they predominantly have mass close to the solar one and use the
results presented after Eq. (\ref{M-J-low}). 
The thermal energy of a solar mass B-bubble taken at the moment
when the Jeans mass dropped down to $M_\odot$ is determined by the
thermal energy of nucleons and electrons, $E_{th} = 3\,T$. Taking 
$T= 25 $ keV, see Eq. (\ref{T-msun}), we find for the total energy
stored inside the B-ball
\be
E_{therm}^{(tot)} = 3T M_B / m_N \approx 1.5\cdot 10^{29} {\rm g} 
\approx  1.5\cdot 10^{50} {\rm erg}\, .
\label{E-therm}
\ee
With the cooling time given by Eq. (\ref{tau-free-num}) the luminosity of
B-bubble would be $L\approx 10^{39}$ erg/sec, i.e. $10^{6}$ of the
solar luminosity.  
Such B-bubbles would be quite bright sources of keV radiation. However,
they would stop radiating their thermal before hydrogen recombination.
The temperature of CMBR at $t=t_{diff}\sim 10^{11}$ s, would
be about 3 eV. If the B-bubbles make all dark energy, i.e. 
$\Omega_{BB} = 0.25$, their mass density at $t_{diff}$ would be
equal to that of CMBR. Correspondingly the energy density 
of the emitted keV radiation would be at most $10^{-4}-10^{-5}$ of CMBR.
This radiation would red-shift today by $10^{-4}$ and move to eV
region, i.e. to background light. 

More efficient source of power could be nuclear reactions that might ignite
inside B-balls. In the considered example with $\rho \sim 10^6$ g/cm$^3$
B-ball has the properties similar to those of the core of red giant
at the initial stage of its evolution. The main source of energy under
these conditions would be helium-4 burning, $3\,^4He \rar ^{12}C$. 
However, the temperature would be somewhat larger, $T\sim$25 KeV, 
instead of 10 keV. 
Since the probability of the above reaction is proportional to $T^{40}$,
the life-time of such helium flash would be extremely short. Naively taken 
these numbers, we obtain life-time about 10 s. However, this simple estimate
can be wrong by several orders of magnitude because the efficiency of the
process is very much different from that in normal giant star. Still 
helium would be burnt very quickly and later nuclear reactions would be
presumably insignificant. More accurate estimates would demand development
of astrophysics for such unusual objects as B-balls and this is not a 
subject of this work. In the following sections we consider observational
manifestations of different types of B-balls allowing for existence of any
type of such objects. Here we note only that if the nuclear physics of B-balls
with the chosen above parameters is similar to that of the giant star core,
the result of helium flash would be a compact star with mass density of
about $10^4$ g/cm$^3$. The flash should take place at relatively early stage
of cosmological expansion
to be safe from the CMBR restrictions. Indeed that relative
energy density of  B-balls would be 
$\rho_{BB}/\rho_\gamma < 10^4/z$. Since nuclear reactions result in emission of
approximately $10^{-3}$ of the total mas of a star, and we do not allow than
more $10^{-4}$ distortion of CMBR, we conclude that such B-balls should consume
their helium at cosmological red-shift $ z > 10^{5}$, i.e. at $T\approx 20$ eV
and $t< 3\cdot 10^9$ s, which does not look unreasonable in view of the
presented above estimate of the life-time $\sim 10$ s.

KeV photons coming from the cooling of B-balls would heat up cosmic
electrons and they in turn might distort the spectrum of CMBR. However,
since the
total energy influx is about $10^{-4}-10^{-5}$ of the total energy density
of CMBR, the spectrum distortion should be not more than at the same level.
More detailed calculations are in order.

If B-bubbles consist of antimatter, the annihilation of the
background matter on their surface would create an additional 
radiation. However, due to a strong coupling of protons and electrons
to photons before hydrogen recombination, the proton diffusion to an
anti-B-bubble is quite slow and it is easy to see that the process
is not efficient. After recombination the number of annihilations
on the surface of one B-ball per unit time would be 
\be
\dot N = 4\pi R_B^2 \beta n_\gamma V_p \approx
10^{31} V_p \left(\frac{T}{ 0.1\,\,{\rm eV}}\right)^3
\left(\frac{R_B}{10^9\,\,{\rm cm}}\right)^2,
\label{dot-N-recomb}
\ee
where $\beta = n_B/n_\gamma = 6\cdot 10^{-10}$ and
$V_p\sim 10^{-5} $ is the hydrogen velocity. 
Assuming that the surface annihilation lasted approximately for one
the Hubble time at $T\sim 0.1$ eV, i.e. $t=t_H \approx 10^{13}$ s
and that the mass density of B-bubbles is not larger than
$10 \, \rho_{CMBR}$, we obtain that the energy density of gamma radiation
from the annihilation on the surface is smaller than $10^{-16}$
of that of CMBR.

A more detailed analysis is of course necessary but it does not fit
the frames of this letter. The dynamics and astrophysics of
B-bubbles in the early universe will be considered elsewhere.
There could be quite interesting signatures of B-bubbles as e.g.
gamma ray background and background light. They might also have an
observable impact on CMBR. We will not go into these interesting
phenomena here. We only want to demonstrate that 
the model of Ref.~\cite{ad-silk}
allows an early production of dark stellar-mass objects consisting
of matter and antimatter which may survive to the present time. 
Such objects would behave as the usual cold dark matter and so they
would populate galactic halos in contrast to the normal stars which
live in the smaller luminous parts of galaxies.

\section{Antimatter in contemporary universe \label{s-late}}
 
In what follows we will not dwell on possible scenarios
of antimatter creation but simply consider phenomenological
consequences of their existence in the present day universe,
in particular in the Galaxy, allowing for any types of such
objects. According to the discussion in the previous section,
astronomically interesting domains of antimatter probably
populate galactic halos but we will not confine ourselves to
this case only, but consider phenomenology of more general
situation. If high density regions of antimatter have survived up 
to the present time, we assume that astronomical anti-objects 
can be either inside and/or in the halo of our galaxy. 
In particular there could be anti-clouds, anti-stars, 
anti-stellar clusters and anti-black holes (i.e. black holes 
generated by the gravitational collapse of antimatter; they 
may be distinguishable from ``standard'' black holes if they 
were surrounded by an anti-atmosphere). Of course, the 
presence of anti-objects in the Galaxy today should lead to 
the production of the gamma radiation from matter-antimatter 
annihilation. Hence we would expect $\sim 100$ MeV $\gamma$ from the 
decay of $\pi^0$-mesons produced in $p \bar p$-annihilation,
with an average of four $\gamma$ per annihilation,
and  $2\gamma$ from $e^+e^-$-annihilation
with $E=0.511 $ MeV if $e^+ e^-$ annihilate at rest. 
In addition to the slow background positrons
there should be also energetic secondary positrons produced by 
pion decays from $\bar p p$-annihilation. Astronomical observations
are seemingly more sensitive to $\bar p p$-annihilation because
the total energy release in $\bar p p$-annihilation is 3 orders of 
magnitude larger than that in $e^+e^-$-annihilation and
the galactic gamma ray background at 100 MeV is several orders of 
magnitude lower than the one at 0.5 MeV. On the other hand,
$e^+e^-$-annihilation gives the well defined line 
which is easy to identify.

In order to study the observational constraints on the
number of such anti-objects and to speculate on possible
future investigations, it is helpful to distinguish two
different regimes of matter-antimatter annihilation,
depending on the ratio $R/\lambda_{free}$, where $R$ is the
size of the anti-object and
\be
\lambda_{free} = {1}/({\sigma_{ann} \, n_{\bar p}})
\ee
is the proton or electron mean free path inside an anti-object, 
with $\sigma_{ann}$ and $n_{\bar p}$ being respectively the 
annihilation cross section for $\bar p p$ or $e^+e^-$ (they
have similar order of magnitude) and the antiproton number 
density in the anti-object.

\subsection{Volume annihilation \label{ss-volume-ann}} 

Volume annihilation takes place if $\lambda_{free}$ is 
larger than the radius $R_B$, that is:
\be
R_B \sigma_{ann} n_{\bar p} \approx
3\cdot 10^{-2} \left(\frac{R_B}{0.1\,{\rm pc}}\right)\,
\left(\frac{n_{\bar p}}{10^4\,{\rm cm}^{-3}}\right)\,
\left(\frac{\sigma }{10^{-23}\,{\rm cm}^2}\right) < 1 \; .
\label{R-sigma-n}
\ee
One should remember that $\sigma_{ann} \sim 1/v$ where $v$ 
is the velocity of the annihilating particles in their 
center-of-mass frame and thus the mean-free path is different 
for protons with different velocities. We assume that the 
typical velocities are of the order of $10^{-3}c$. The effective
time of annihilation, $\tau = \lambda_{free}/ v$ does not depend on velocity
of the annihilating particles.

The annihilation rate per unit time and volume is equal to
\be 
\dot n_{\bar p} = \sigma_{ann} v n_p n_{\bar p} \; ,
\label{dot-bar-n}
\ee
where $n_p\approx 1/{\rm cm}^3$ is the average galactic number 
density of protons. Hence the life-time of ``volume-annihilated'' 
objects would be
\be
\tau_{vol} = \left(\sigma_{ann} v n_p \right)^{-1} \approx
10^{15}\,{\rm s}\, \left( \frac{ {\rm cm}^{-3}}{n_p}\right)\,
\left(\frac{10^{-15}{\rm cm}^3/s}{\sigma_{ann} v}\right) \; .
\label{tau-vol}
\ee
The result does not depend upon the size and density of the
anti-object.

Thus the low density anti-objects which are annihilated by 
surrounding protons inside their whole volume could not survive 
in galaxies to the present time, $t_U \approx 3\cdot 10^{17}$ s.
However, as we argued in Sec. \ref{s-evolution}, B-balls could
naturally populate  galactic halos, where the proton number 
density is lower, about $10^{-4}/{\rm cm}^3$, and the 
life-time of the anti-object 
would be at the level of $\tau_{vol} \approx 10^{19}$ s.

The luminosity in gamma-rays of the volume-annihilated objects 
can be estimated as follows. The total energy release per unit 
of time is given by $L_{tot}^{(vol)} = 2 m_p \dot N_{ann}$, where 
$m_p$ is the proton mass. The luminosity in gamma rays, produced 
by the annihilation, would be
\be
L_\gamma^{(vol)} \approx 0.3 L_{tot}^{(vol)} \approx
 10^{35}\,{\rm erg/s}\, 
\left( \frac{\sigma_{ann}v}{10^{-15}\,{\rm cm^3/s}}\right)
 \left(\frac{R_B}{0.1\,{\rm pc}}\right)^3
 \left( \frac{n_p}{{\rm 10^{-4}\,cm}^{-3}}\right)
 \left(\frac{n_{\bar p}}{10^4 {\rm cm}^{-3}}\right) \; ,
\label{L-gamma-vol2}
\ee
if the mean free path of photons inside the anti-object is larger
than its size, $R_B$. If such an object is at the 
distance of 10 kpc from the Earth, the expected flux would be about 
$10^{-11}$ erg/s/cm$^2$, which corresponds to  $10^{-7}$ photons/s/cm$^2$.
It is below the gamma ray background 
with energy $\sim$ 100 MeV and in the sensitivity range of
the present and future experiments.

We would expect approximately of the same photon 
flux from $e^+e^-$-annihilation 
because $\sigma v$ for $e^+e^-$ is twice bigger than that for $\bar p p$, 
but number of photons per annihilation is twice smaller. Thus the
flux of 0.511 MeV line from an anticloud at 10 kpc 
would be about $10^{-7}$ photons/cm$^2$/s
and it should be compared with the observed flux of the same line
equal to $\sim 10^{-4} $ photons/cm$^2$/s~\cite{e-line}.

Such anticlouds could also be observed in visible light, created
by the bremsstrahlung of the energetic electrons originated from
$\bar p p$-annihilation. The probability of such processes is
two orders of magnitude smaller than of the main one. The energy
release is roughly 8 order of magnitude smaller. So the visible
luminosity would be 8 orders of magnitude weaker than the solar one.

The influx of protons could be diminished by the radiation pressure, 
allowing the  anti-objects to survive up to the present time in 
more dense regions as well. This effect is considered in 
Subsection~\ref{ss-eddi}, where we see, however, that the radiation pressure
is usually not essential.

\subsection{Surface annihilation \label{ss-suface-ann}}

If the proton mean free path is much smaller than the size
of the anti-object, $\lambda_{free} \ll R,$
the annihilation takes place on the surface. This is 
typical situation for stellar types anti-objects and,
even more, for compact anti-stars, as e.g. white
dwarfs, (anti)-neutron stars, etc.
In this case all the protons that hit the surface of the 
anti-object annihilate. The annihilation cross section is 
given by the geometrical area of the anti-object,
$\sigma = 4 \pi R^2$, and the gamma-ray luminosity of such 
a compact anti-object is equal to:
\be
L_\gamma^{(sur)} \approx 4\pi R_B^2\, 0.6 m_p F_p
\approx 10^{27}\,{\rm erg/s}\,\left( \frac{R_B}{R_\odot} \right)^2
\left(\frac{n_p}{{\rm cm}^{-3}}\right) \left(\frac{v}{10^{-3}}\right) \; ,
\label{L-sur}
\ee
where $F \sim n_p v$ is the proton flux and
$R_\odot \sim 7 \cdot 10^{10}$ cm is the Solar radius.
With this luminosity a solar mass anti-star would have 
the life-time of the
order of $10^{27}$ s, if all the factors in Eq. (\ref{L-sur})
are of order unity.

If such an anti-star lives in the galactic center, 
where $n_p \gg 1/$cm$^3$, its 
luminosity may be quite large. However, the pressure of
emitted gamma radiation could reduce the proton flux and
diminish $L_\gamma^{(sur)} $. This effect is considered in
the following subsection and we see that quite high
luminosities are permitted.

\subsection{Eddington limit \label{ss-eddi}}

If the luminosity of an object is created by the influx of
particles, there exists an upper bound on its
luminosity, which follows from the fact that the pressure of the
emitted radiation diminishes the incoming flux.

The force acting on proton, or time-derivative of the proton
momentum, $P$, created by the gamma ray pressure at distance
$r$ from the radiating objects is equal to:
\be 
\dot P \sim \sigma_{p\gamma} n_\gamma (r) \omega \; ,
\label{dot-P}
\ee
where $\sigma_{p\gamma}\sim 10^{-31} $ cm$^2$
is the cross-section of the Compton
scattering on proton, $\omega$ is the gamma ray energy and
$n_\gamma$ is the number density of emitted photons. The total
luminosity of the object is  
$L= 4\pi R^2 n_\gamma (R) \omega$. Here $n_\gamma$ is taken at
the surface of the object, but it drops as $(R/r)^2$ with the
increasing distance $r$.

The force of the gravitational attraction acting on the protons
at distance $r$ is equal to
\be
F_{grav} = \frac{G_N m_p M}{r^2} \; .
\label{F-grav}
\ee
Demanding that the gravitational attraction should be stronger
than the radiation pressure, we obtain: 
\be
L< 10^{45}\,{\rm erg/s}\, \left(\frac{M_B}{M_\odot}\right)
\left(\frac{10^{-31}\,{\rm cm}^2}{\sigma_{p\gamma}}\right) \; .
\label{edd-limit}
\ee

One may argue that the total flux of the particles to the anti-object
must be electrically neutral and hence the much larger force
exerted by photons on electrons should be substituted into
Eq. (\ref{dot-P}). For the low energy photons, $\omega <m_e$, 
the Thomson cross-section, $\sigma_{Th} = 0.66\cdot 10^{-24}\,{\rm cm}^2$,
should be used. In the case of photons originating from the 
$\bar p p$-annihilation, their energies are much larger and the
cross-section is about $\sigma_{e\gamma} \sim \pi\alpha^2/(m_e\omega)$.
This would diminish bound (\ref{edd-limit}) by roughly 4 orders of
magnitude. In any case, the Eddington limit is well above 
prediction (\ref{L-sur}).

On the other hand, the excessive charge created by a larger
influx of protons with respect to electrons
could be compensated by the out-flux of positrons from the
antimatter object. If this is the case, the limit on the
gamma-ray luminosity would be given by Eq. (\ref{edd-limit}).
Energetic gamma rays from $\bar p p$-annihilation would be accompanied by 
the emission of low energy positrons, which would be a source of
0.511 MeV line.

\section{Point-like sources of gamma radiation}
\label{s-sources}

First, we consider the possibility of presence of 
anti-clouds in our galaxy. If condition (\ref{R-sigma-n}) is
satisfied, the proton mean free path
inside the anti-cloud is larger than the anti-cloud size and
matter-antimatter annihilation proceeds in the whole
volume. The life-time of such clouds is given by Eq. (\ref{tau-vol}),
if the proton flux is sufficiently large. Such clouds would not
survive to the present time. However, if the proton flux is
not big enough, the life-time of the cloud may exceed the age
of the Galaxy. 

More favorable condition for survival of anti-clouds are in the
galactic halo, where the proton density is 
$n_p \sim 10^{-4}$ cm$^{-3}$. The gamma-ray luminosity
of such a cloud is given by Eq. (\ref{L-gamma-vol2}). 
Such a source might be observed on the Earth as
a $\gamma$ ray source with the flux:
\be\label{anti-cloud}
\phi_{Earth} \sim  10^{-7}
 \left( \frac{n_p}{10^{-4}\,{\rm cm}^{-3}}\right)
 \left( \frac{n_{\bar p}}{10^{4}\,{\rm cm}^{-3}}\right)
\left(\frac{R_B}{0.1\,{\rm pc}}\right)^3
\Big(\frac{10 \, \rm kpc}{d}\Big)^2\;
{\rm cm^{-2} \; s^{-1}} \; ,
\ee
where $d$ is the distance of the anti-cloud from the Earth.
Eq. (\ref{anti-cloud}) should be compared with the point
source sensitivity of EGRET \cite{egret}, at the level of 
10$^{-7}$ photons cm$^{-2}$ s$^{-1}$ for $E_\gamma > 100$ MeV
and a full two weeks exposure, and of the near-future 
GLAST \cite{glast}, which is about two order of magnitude 
better, $\sim 10^{-9}$ photons cm$^{-2}$ s$^{-1}$.

As for possible anti-stars, emitting gamma rays from 
$p\bar p$-annihilation on their surface,
they should be quite close to us
in order to be detectable point-like sources. 
For an anti-star in the galactic disc,
its $\gamma$ flux would be
\be\label{flux-anti-star}
\phi_{Earth} \sim 10^{-7}
\Big(\frac{R_B}{R_\odot}\Big)^2
\Big(\frac{1 \, \rm pc}{d}\Big)^2 \; 
{\rm cm^{-2} \; s^{-1}} \; .
\ee
To be observable in a near future such an anti-star should be 
really in solar neighborhood and if it is a normal star
powered by thermonuclear energy it could even be seen with a 
naked eye. On the other hand, if anti-stars are similar to the 
Sun, the consideration of a possible anti-stellar wind requires
that their number in the Galaxy is smaller than $10^5$ 
(see below Eq. (\ref{bound-wind})) and, assuming
they are uniformly distributed in the galactic disk, we would 
expect a mean anti-star number density of $1/(100 \; {\rm pc})^3$.
In this case $d \sim 100$ pc implies $\phi_{Earth} \sim 10^{-11}$
cm$^{-2}$ s$^{-1}$, a photon flux too weak to be detectable from 
a point-like source.

The $\gamma$-flux from an anti-star may be strongly enhanced if 
it happens to be in a high density hydrogen cloud. This 
possibility is quite realistic: even if the mean proton density
is $n_p \sim 1$ cm$^{-3}$, about one half of the interstellar medium
is tied up in gas clouds with average proton density 
$n_p \sim 10^3$ cm$^{-3}$. In other words, gas clouds occupy
a fraction of about $10^{-3}$ of the Galaxy volume, so that, 
assuming $10^5$ anti-stars in the Galaxy, about 100 of them 
are expected to live in some gas cloud. Their gamma flux would 
increase by three orders of magnitude, or even more if we 
consider that some clouds are much denser, up to 
$n_p \sim 10^6 - 10^9$ cm$^{-3}$.

Last, it could be interesting to consider observational
signatures of anti-black holes, i.e. black holes generated
by the gravitational collapse of antimatter. They may be distinguished
from the ordinary black holes if they were surrounded by an atmosphere
of antimatter (it is not difficult to imagine that it is 
possible, even if detailed calculations would be necessary for a precise
assertion). Their anti-atmosphere could be considered as an anti-cloud 
around a very compact anti-object and, following the considerations
of Subsection \ref{ss-volume-ann}, such an anti-atmosphere could 
survive up to the present day only if such an anti-black hole was 
in the galactic halo. In this case this strange anti-object could 
be detectable looking for high energy $\gamma$ radiation (see 
Eq. (\ref{anti-cloud})) from a stellar mass object creating 
gravitational micro-lensing (MACHO)~\cite{machos}. The same would be 
true also for compact anti-stars in galactic halo.

The spectrum of photons from matter-antimatter annihilations
is well known and consists of three different parts. First, most
energetic are photons from $\bar p p \rar \pi^0 \rar 2\gamma$. 
If $\pi^0$ were at rest we would observe the single 67.5-MeV line. 
However, the life-time of $\pi^0$ is very short and they decay 
being relativistic. Thus the spectrum spreads  both ways up and 
down and shifts from zero to higher energies above $200$ MeV. 
The second less energetic and also continuous part comes from the 
chain of reactions $\bar p p \rar \pi^+ \rar \mu^+  \rar e^+$ and
subsequent $e^+e^- $-annihilation. The probability of such double
annihilation and the shape of the spectrum depends upon the object 
where these processes take place. This process seems to be least
efficient. The annihilation of the slow positrons inside antimatter
objects with the background electrons leads to the famous 
0.511 MeV line, which is easy to identify. This anomalously bright
line is observed
recently in the Galactic center~\cite{gr-an}, Galactic 
bulge~\cite{e-continuum} and possibly even in the 
halo~\cite{e-line}. Though an excess of slow positrons is 
explained in a conventional way as a result of their creation
by light dark matter particles, such a suggestion is rather 
unnatural because it requires a fine-tuning of the mass of the dark 
matter particle and the electron mass. More natural explanation is
the origin of these positrons from primordial antimatter objects.

\section{Diffuse galactic gamma ray background}
\label{s-background}

In the standard theory, the galactic production of $\gamma$
rays is due to inelastic collisions of high energy cosmic 
rays with the interstellar medium (the dominant processes
are $p+p$, $p+\alpha$ and $\alpha+\alpha$), to bremsstrahlung
radiation from cosmic ray electrons and to inverse Compton
scattering of electrons with low energy photons. Many authors 
have calculated the galactic production rate of $\gamma$ 
per hydrogen atom (see e.g. Ref. \cite{mori}) and the estimated 
rate in the energy range $E_\gamma > 100$ MeV is
\be\label{standard}
\Gamma_\gamma \sim 2 \cdot 10^{-25} \; {\rm s^{-1} 
\; atom^{-1}} \; ,
\ee
in good agreement with observational data \cite{digel}. From 
Eq. (\ref{standard}) we can deduce the total production rate
of high energy $\gamma$ rays in our galaxy
\be\label{total}
\Gamma_\gamma^{tot} \sim 10^{43} \; {\rm s^{-1}} \; .
\ee

The presence of $N_{\bar S}$ anti-stars in the galactic disc,
where the average proton number density is $n_p \sim 1$ cm$^{-3}$,
would create a new source of high energy $\gamma$ rays, with 
the contribution
\be\label{anti-stars}
\dot{N}_\gamma \sim \dot{N}_{\gamma}^{(sur)} \, N_{\bar S} \; ,
\ee 
where 
$\dot{N}_{\gamma}^{(sur)}\approx L_{\gamma}^{(sur)}/(100\;{\rm MeV})$
is the photon flux coming from each anti-star and  $L_{\gamma}^{(sur)}$ 
is given by Eq. (\ref{L-sur}). If we assume that 
$\dot{N}_\gamma$~(\ref{anti-stars}) 
cannot exceed 10\% of the standard galactic production rate of 
high energy $\gamma$, given by Eq. (\ref{total}), we obtain a bound 
on the present number of anti-stars (for simplicity, we assume that 
all the anti-stars have the same radius $R_B$)
\be
N_{\bar S} \lesssim 10^{12} \Big(\frac{R_\odot}{R_B}\Big)^2 \; .
\label{bound-gamma}
\ee

A stronger constraint can be obtained from the consideration of
the annihilation of antimatter from the anti-stellar wind with
protons in the interstellar medium. The rate of $\bar p$
emission by the anti-stellar wind per anti-star is  
\be
\dot{N}_{\bar p}^{wind} \sim 10^{36} \, W \; {\rm s^{-1}} \; ,
\ee
where $W = \dot{M}/\dot{M}_\odot$ is the ratio of an
anti-star mass loss rate to the Solar one. 
The total number of antiprotons
in the Galaxy can be determined by an equation: 
\be
\dot{N}_{\bar p}^{tot} = S - T \; ,
\ee 
where $S = \dot{N}_{\bar p}^{wind} \, N_{\bar S}$ is the source 
term and $T$ the sink term due to annihilation of $\bar p$. 
The life-time of antiprotons in the Galaxy with respect to 
annihilation is given by Eq. (\ref{tau-vol}), i.e. it is much 
smaller than the age of the Galaxy. Thus the number of 
antiprotons reached a stationary value, i.e. 
$\dot{N}_{\bar p}^{tot} = 0$ and the 
production rate of 100 MeV $\gamma$ in the Galaxy is 
$\dot{N}_\gamma \approx 4 \, S$, because in each act of 
$\bar p p$-annihilation 4 photons are produced  on the average.
The flux of 100 MeV photons on the Earth would be
\be
\phi_{\bar p p}^{(gal)} \approx 
\frac{4 \dot N_{\bar p}^{wind} N_{\bar S}} 
{4 \pi R_{gal}^2} \approx 100 W
\left(\frac{N_{\bar S}}{N_S}\right)\,
{\rm cm}^{-2} {\rm s}^{-1},
\label{phi-gal}
\ee
where we took $N_S= 3\cdot 10^{11}$ and $R_{gal} = 10$ kpc.
The luminosity of the Galaxy in 100 MeV gamma rays 
from anti-stellar wind would be 
$L_{\bar S} \sim 10^{44} W\,N_{\bar S}/N_S$ erg/s.
Since from Eq. (\ref{total}) we find that the total
Galaxy luminosity in 100 MeV $\gamma$ is
$L_{\gamma}^{tot} \sim 10^{39}$ erg/s, 
the related bound on the number of anti-stars is:
\be\label{bound-wind}
N_{\bar S}/N_S \lesssim 10^{-6} \, W^{-1} \, .
\ee
Here, as always we assume that 
the contribution from new physics cannot
exceed 10\% of $L_{\gamma}^{tot}$.

A similar restriction can also be obtained from the 0.511 MeV
line created by $e^+ e^-$-annihilation with positrons from
the anti-stellar wind. Since the number of antiprotons in the
stellar wind is approximately the same as the number of 
positrons, the flux of 0.511 MeV photons would be close to 
that given by Eq. (\ref{phi-gal}). Taking that the latter is
smaller than $10^{-4}$ /cm$^2$/s we find 
\be 
N_{\bar S}/N_S \lesssim 10^{-6} \, W^{-1} \, .
\label{N-bar-S}
\ee

If anti-stars have been formed in the very early universe
in the regions with a high antibaryonic density~\cite{ad-silk}, 
such primordial stars would most probably be compact ones, 
white or brown dwarfs, neutron stars, etc. The stellar
wind in this case would be much smaller that the solar one, 
$W\ll 1$. Their luminosity from the annihilation 
on the surface should be very 
low, because of their small radius $R$, and their number in 
the Galaxy may be as large as the number of the usual stars. 
This possibility is not excluded by the bounds (\ref{bound-gamma}) 
and (\ref{bound-wind}). 
Such compact dark stars could make a noticeable part of the 
cosmological dark matter. As we have argued in Sec. \ref{s-evolution},
the early created compact stellar like objects behave as 
the usual collisionless cold dark matter (CDM). In this case it is 
natural to expect that they would be distributed in and around 
galaxies as the standard CDM, having a large number density in
the galactic center and decreasing as $1/r^2$ in the halo.

These compact objects
would generate the diffuse gamma-ray background not only now
but during all cosmological history. In particular, we expect
a strong constraint on the number of anti-stars from
their emitted radiation during the so called ``dark age'', after
recombination but before the advent of the early standard stars.
We thank Francesco Villante for having pointed out this 
possibility. We plan to evaluate the intensity of such 
cosmological $\gamma$-background in another work.

\section{Anti-nuclei in cosmic rays}
\label{s-cr}

Stable charged particles in cosmic rays consist of 86\% of 
protons, 11\% helium nuclei, 1\% heavier nuclei, and 2\% 
electrons. It is common belief that the abundances of the 
elements in the cosmic rays reflect relative abundances in 
the Galaxy (even if the low energy cosmic rays should be a mirror 
of the relative abundances in the Solar System). Hence, we 
can reasonably expect that the antimatter-matter ratio in 
cosmic rays is more or less equal to the (anti-star)-star 
ratio $N_{\bar S}/N_S$, if the antistars are of the same kind
as the stars in the Galaxy. As for antiprotons and positrons, 
they are not direct indicators for the existence of primordial
antimatter, because they can be produced in many 
astrophysical processes. For example, the observed ${\bar p}/p$ 
ratio is at the level of 10$^{-4}$ and is compatible with 
theoretical predictions for $\bar p$ production by the high energy 
cosmic ray collisions with the interstellar medium. A possible 
contribution of $\bar p$ from exotic sources (ES) is not more 
than about 10\% of the total observed $\bar p$ flux, so the 
number of anti-stars $N_{\bar S}$ have to be
\be
\frac{N_{\bar S}}{N_S} \lesssim \left(\frac{\bar p}{p}
\right)_{\rm ES} \lesssim 10^{-5}
\quad \Rightarrow \quad N_{\bar S} \lesssim 10^6 \; ,
\ee
since the number of ordinary stars in the Galaxy is 
$N_S \sim 10^{11}$.

On the other hand, the possibility of producing heavier 
anti-nuclei (such as anti-helium) in cosmic ray collisions is 
completely negligible and a possible future detection of the 
latter would be a clear signature of antimatter objects. At 
present there exists an upper limit on the anti-helium to helium 
ratio in cosmic rays, at the level of 10$^{-6}$ \cite{bess}, 
leading to the constraint
\be
\frac{N_{\bar S}}{N_S} \lesssim \left(\frac{\bar{He}}{He}
\right)_{\rm ES} \lesssim 10^{-6}
\quad \Rightarrow \quad N_{\bar S} \lesssim 10^5 \; ,
\ee
which is essentially equal to that from anti-stellar wind in 
Eq. (\ref{bound-wind}). However, the sensitivity of AMS \cite{ams} 
and PAMELA \cite{pamela} space missions is expected to be two 
orders of magnitude better, at about 10$^{-8}$. In this case 
the non-observation of anti-helium nuclei would lead to the much 
stronger constraint 
\be\label{bound-ams-pam}
N_{\bar S} \lesssim 10^3 \; .
\ee
Of course, the bound (\ref{bound-ams-pam}) cannot be applied if
anti-stars are compact ones from the very beginning (i.e. from the 
moment of their formation in the early universe). In this case the 
stellar wind from them and the shortage of anti-supernova events
would spread much less (anti)helium than the normal stars,
but heavier anti-elements might be not so strongly suppressed.
According to the scenario described in Sec. \ref{s-evolution},
the compact (anti-)stars might form in (anti)baryon-rich regions, 
where the primordial nuclear abundances would be quite different 
from the standard ones with an enhanced amount of heavier (anti-)nuclei, 
as e.g. oxygen, nitrogen, carbon, and maybe even heavier one up to 
calcium or iron~\cite{ibbn}. Spectroscopy observations may detect 
stellar atmospheres anomalously rich in heavy elements. Such chemical 
anomaly is a good signature that such object are made of antimatter 
and a search for gamma rays from them looks promising. On the other 
hand, the mechanisms capable of ejecting antimatter from such stars 
and of spreading it out in the Galaxy are favorable for enrichment 
of galactic cosmic rays with heavier anti-nuclei. Unfortunately the 
amount of anti-nuclei depends upon many unknowns and it is impossible 
to make a reliable estimate of their flux.

\section{More exotic events}
\label{s-spectacular}

The presence of anti-stars in the Galaxy could lead to
extraordinary events of
star-antistar annihilation. As a matter of fact, the
radiation pressure produced in the collision prevents
their total destruction. Still the released energy can be 
huge.

The most spectacular phenomenon is a
collision between a star and an anti-star with similar
masses $M$. 
The energy released in such a collision can be estimated as
follows. The relative momentum of the colliding stars is
approximately  $P\sim M v$, where the typical value of the relative
velocity is about $v\approx 10^{-3}$. This momentum would be
``canceled'' by the pressure of radiation created by 
baryon-antibaryon annihilation, $p$. The total pressure force 
is $F\sim \pi r^2 p$, where $r$ is the radius of the circle
where the colliding stars penetrated into each other.  
The bounce would take place when $F t_{coll} \sim Mv$, where
$t_{coll} \sim d/v$ is the collision time and $d$ is the 
penetration depth. Since the annihilation products are relativistic,
their pressure density is of the same order of magnitude as the
energy density and hence the amount of the annihilated matter
during the star-antistar collision would be 
\be
m_{ann} \sim Mv^2 \; ,
\label{m-ann}
\ee
i.e. the total energy release would be 
\be
\Delta E_1 \sim 10^{48}\, {\rm erg} (M/M_\odot) (v/10^{-3})^2 \; . 
\label{Delta-E}
\ee
Most probably the radiation would be emitted in a narrow disk 
along the boundary of the colliding stars.

The collision time can be estimated as follows. Let us introduce the
angle $\theta$ at which the radius, $r$, of the collision disk is seen
from the star center, $r = R\theta$, assuming that
$\theta <1$. Here $R$ is the star radius.
The penetration depth is $d= R \theta^2$. The volume where matter and
antimatter mix and annihilate is $R^3\theta^4$. This volume
should be about $v^2 $ of the total stellar
volume. Thus the penetration depth is $d \sim vR $ and the
collision time is $t_{coll} \sim R $. For the solar type star
this time is about 3 s. For colliding compact star-antistar the
collision time would be much smaller but the energy release could be
much larger, because the velocity might approach relativistic
values.

We expect that in the process of a star collision with 
the similar anti-star the energy would be emitted inside a 
narrow disk with the opening angle $\theta \sim \sqrt{v}$. 
The characteristic time of the emission is of the
order of a second. The energy of the radiation should be 
noticeably smaller than 100 MeV because the radiation
should degrade in the process of forcing the star bounce. 
This makes this collision similar to gamma bursts but 
unfortunately some other features do not fit so well:
the released energy should be much larger, 
about $10^{53}\,\sqrt{v}$~erg and it is
difficult to explain the features of the afterglow. A 
reasonable amount of the energy could be released in the
case of compact star annihilation. 
It may be possible to explain two or more bursts by
the oscillating motion with interchangingly dominated
attraction by gravity and repulsion
by the pressure of the products of annihilation. The process
is surely more complicated than our naive picture and we
cannot exclude that gamma bursts are the results of 
star-antistar annihilation.

The collision of a compact star, e.g. an anti-neutron star,  
with a usual one or with a red giant would look completely 
different. In this case, the mass densities of the two
objects are so much diverse that the anti-neutron star
would go through the red giant without stopping and would 
annihilate all what it meets on the way.
The released energy is about 
\be
\Delta E_2 \sim  \pi R^2_{ns}\, D\, \rho_{rg} = 
M_{rg} \left(\frac{R_{ns}}{R_{rg}}\right)^2
\left(\frac{D}{R_{rg}}\right) \; ,
\label{E-2}
\ee
where $R_{ns}$ is the radius of the anti-neutron star,
$D$ is the crossed distance inside the red giant, $R_{rg}$
is the radius of the red giant, 
$\rho_{rg}$ is the red giant mass density, and $M_{rg}$ is its mass.
For $M_{rg} \sim M_\odot$ and $R_{rg} \sim 10^{14} $ cm
a reasonable estimate of this energy release is 
$10^{38}$ erg. The crossing time of the red giant is about 
$R_{rg}/v\sim  3\cdot 10^6$ seconds. Hence the additional 
luminosity during anti-neutron star propagation inside the red
giant would be an order of magnitude smaller than the  solar 
luminosity and, probably, most of the energy would reheat the
interior of the star and could not reach the free space. 
These estimates are valid for collision of a compact antistar
with the envelope of red giant. A collision of antistar with
the core of red giant would look similar to the compact
star-antistar collision considered above. Most spectacular would
be a collision of red-giant with anti-red-giant, where a fantastic
amount of energy would be released.

The probability of the collision of two stars 
can be estimated as follows. 
The collision rate is $\Gamma = \sigma\, n_S\, v$, where 
$\sigma \approx \pi R^2$ is the geometrical area of the
larger star and $n_S \sim 1/{\rm pc}^3$ the mean star number
density in the Galaxy. The
total number of collisions per unit time would be:
\be
\dot{N}_{\bar S} = \Gamma N_{\bar S} =
\pi v R^2 N_{\bar S} \, n_S 
\approx 10^{13}\,{\rm year}^{-1}\, \left(\frac{N_{\bar S}}{10^5}\right)
\left(\frac{R}{10^{11}\,{\rm cm}}\right)^2 \; .
\label{dot-N-barS}
\ee
If anti-stars are similar to the Sun, from the bound 
(\ref{bound-wind}) we find $N_{\bar S} \lesssim 10^5$, which 
implies essentially no collisions during the whole history 
of the universe. On the other hand, if the anti-stars are 
white dwarfs or anti-neutron stars, the bounds (\ref{bound-gamma}) 
and (\ref{bound-wind}) are very weak or inapplicable and 
the anti-star number can be as large as that of the ordinary 
stars, i.e. $N_{\bar S} \sim 10^{11}$. In this case we find 
one collision per $10^7$ years.

The collision with a red giant would have a larger cross-section 
because of much larger radius of the red giant, $R_{rg} > 10^{13}$ cm.
Since red giants are about 1\% of the stars in the Galaxy,
the number of collisions could increase by 2 orders 
of magnitude. Another factor which might also 
enhance the probability is a larger 
gravitational attraction of heavy stars.
We know, the majority of the stars 
in the Galaxy are in multiple stellar systems. However, it is
surely true for the normal stars which were formed from the
primeval hydrogen cloud in galactic disk, while B-balls  
were created in the early universe by a completely
different mechanism. However, the gravitational capture of a 
normal star and B-ball has the same probability as the gravitational
capture of the normal stars, see discussion in the next paragraph.

Another interesting possibility is the transfer of
material in a binary system. In the case of the ordinary stars
made of matter, it can happen that a binary system is
formed by a red giant with a close compact star companion
such as a white dwarf. If the former overflows the Roche
lobe of the latter, the compact star captures gases from
the red giant outer atmosphere. Then, on the white dwarf
surface, a large amount of hydrogen is rapidly
converted into heavier elements via the CNO cycle,
producing an extremely bright outburst of light.
The event taking place can be a nova, where the star luminosity
at the brightness peak is about 10$^{38}$ erg/s, or the
much more spectacular phenomenon of supernova Ia, with a 
maximum luminosity of about 10$^{52}$ erg/s (however, the latter 
case is not of interest for us, because the emitted energy comes
from star gravitational collapse and does not from the CNO cycle).
We can reasonably expect that something similar can
happen in a binary system where one of the stars is made
of matter and the other one of antimatter. In this case 
hydrogen is not burnt via the CNO cycle, but the much more 
violent process of matter-antimatter annihilation takes 
place and the white dwarf should be brighter than an 
ordinary nova. Of course, the two stars 
cannot be born at the same time from the same cloud, but
they must have a different origin. For example, one star
can be gravitationally captured by another which is already
in a multiple star system (energy conservation forbids the
formation of bound systems from two stars) and, even if today it
is ruled out as the dominant formation mechanism, it is
expected to be not rare (see e.g. Ref. \cite{grav-capt}).
The event could be observed as a long outburst of 100 MeV
$\gamma$ radiation and be easily distinguished from the
standard novae: the star luminosity could not exceed 
the Eddington limit~(\ref{edd-limit}), but 
would be probably close to it. Moreover, even if
the rate of the ordinary novae in the Galaxy is estimated to be 
only about 50 events per year, thanks to the large amount 
of energy release, we can monitor many galaxies at the same 
time, even quite distant from us, and increase our 
possibility of observing this kind of phenomena. Even in this
case, clear predictions are impossible, because observational
probabilities depend on many unknowns.

\section{Conclusion}
\label{s-conclusion}

The conclusion to this paper, unfortunately, is not conclusive -- 
practically anything is allowed. Gamma rays from $\bar p p$-annihilation
may be observable with future or even with existing $\gamma$-telescopes.
Quite promising for discovery of cosmic antimatter are point-like
sources of gamma radiation. 
The problem is to identify a source which is suspicious to consist
of antimatter. A possible manifestation of such a source is an anomalous
abundance of chemical (anti-)elements around it, which can be measured
by spectroscopy.

The 100-MeV gamma ray background does not have pronounced features 
which would unambiguously tell that the photons came 
from the annihilation of antimatter.
The photons produced as a result of $\bar p p$
annihilation would have a well known spectrum but it may be
difficult to establish a small variation of the conventional spectrum
due to such photons.

In contrast, the 0.511 MeV line must originate from $e^+e^-$ 
annihilation and it is tempting to conclude that the observed 
excessive signal
from the Galaxy and, especially, from the galactic bulge come from
astronomical antimatter objects.

An important feature of the scenario of early formed compact antimatter
objects~\cite{ad-silk} is that such stellar type ones would behave
as normal cold dark matter and they would concentrate in or around
normal galaxies.

Interesting candidates for being anti-matter stars are the observed
MACHO events~\cite{machos}. According to the scenario of Ref.~\cite{ad-silk},
these stellar mass gravitational lenses 
should consist of practically equal number of
matter and antimatter objects. The latter should emit 
100 MeV and 0.511 MeV gamma rays and, though the 
luminosity might be rather weak, see Eq. (\ref{L-sur}),
they still may be observed with high angular resolution telescopes. The 
gamma ray luminosity would be much stronger for MACHOs in the
galactic disk, because the number density of protons there is at least 4
orders of magnitude higher than in the galactic halo.

If an antistar happens to be in the Galactic Center, its luminosity
from the surface annihilation of the background matter should be
strongly enhanced due to the much larger density 
of the interstellar matter
there. So the search of the antimatter signatures in
the direction of the Center is quite promising.

There is a non-negligible chance to detect cosmic anti-nuclei
and not only light anti-helium but also much heavier ones,
especially if anti-stars became early supernovae.

A possible discovery of cosmic antimatter would shed light on the
mechanism of baryogenesis and CP-violation in cosmology. 
In the standard scenarios the baryon asymmetry $\beta$ is 
constant, just one number, and it is impossible to distinguish
between different mechanisms of baryogenesis measuring this single
number. More exotic models predicting a noticeable amount of
antimatter in our neighborhood are much more interesting because
$\beta = \beta (x)$ is a function of space points and contains much
more information about physics in the early universe.

Macroscopically large pieces of antimatter not far form the 
Earth may be interesting energy sources, but this is not in
foreseeable future, maybe only in science fiction.

\section*{Acknowledgments}

We thank Filippo Frontera and Francesco Villante 
for useful comments and suggestions.

\end{document}